\documentclass[aps,prb,twocolumn]{revtex4}
\usepackage{graphicx}% Include figure files
\usepackage{dcolumn}% Align table columns on decimal point

\begin{document}

\title{Low-energy Spectra of the $t$-$J$-Type Models Studied by Variational Approach}

\author{Chung-Pin Chou$^{1,2}$, T. K. Lee$^{2}$, and Chang-Ming Ho$^{3}$}
\address{$^{1}$Department of Physics, National Tsinghua University,
Hsinchu 300, Taiwan\\
$^{2}$Institute of Physics, Academia Sinica, Nankang 115, Taiwan\\
$^{3}$Department of Physics, Tamkang University, Taipei 251, Taiwan}

\date{\today}

\begin{abstract}
We discuss recent progress of understanding the phenomena observed in high $T_{c}$ cuprates by studying 
the $d$-wave resonating-valence-bond ($d$-RVB) based trial wave functions for the 2-dimensional $t$-$J$-type models. 
Treat exactly the strong correlation between electrons by numerical approach, we examine 
the evolution of ground states described by different variational wave functions and properties of the 
quasi-hole and -particle excitations of the $d$-RVB superconducting (SC) state. Properties related to the Fermi 
surface geometry deduced from quasi-hole energy dispersion of the SC state is shown to be consistent with 
the observation by photoemission spectroscopy.  
With the calculated spectral weights (SW's) for adding and removing an electron, we found not only the anti-correlation 
between conductance peak height and width between peaks seen in tunneling experiments, but also unique properties 
due to strong correlation which need to be verified by future experiments. 
\end{abstract}

\maketitle

The emergence of superconductivity as holes or electrons doped 
into the insulating parent compounds is one of the fascinating facts 
seen in high $T_c$ cuprates \cite{bonn2006}. In addition to that, there are 
many other intriguing phenomena observed in the underdoped regime which may 
indicate new and even exotic phases \cite{bonn2006}. Phases at different 
doping levels have been 
characterized experimentally by studying in detail 
the low-energy, single-particle spectra by applying, for example, angle-resolved 
photoemission spectroscopy (ARPES) and scanning tunneling 
microscopy/spectroscopy (STM/STS) \cite{arpes,stm0,stm1,fang,stm2}. 

To understand these phenomena, we have been studying the trial wave 
functions (TWF's), in the form of the resonating-valence-bond (RVB) state 
proposed by Anderson \cite{pwa87}, for the $t$-$J$-type model with different 
concentration of doped holes or electrons. Some recent progress 
of this work is discussed in this paper. We start by discussing the different variational WF's 
describing ground states at different doping levels. Although our study is for 
only the uniform, 2-dimensional case 
representing the ideal $CuO_2$ plane without any disorders (due 
to lattice, inhomogeneous doping {\it etc.}), the phases we found, in particular, the 
antiferromagnetic metal (AFMM) \cite{lhn} and the coexisting state of superconductivity and 
antiferromagnetism (AFMM+SC) \cite{ltp} (see Fig.{\ref{phase-diagram}} below)- have recently been 
observed, for the first time, in the Hg based 5-layer cuprates \cite{NMR}. 

We have also extended 
the TWF approach to examine properties of the low-energy excitation of 
the $d$-wave RVB ($d$-RVB) state in the superconducting regime. In the second part of this 
paper, we present the doping dependence of Fermi surfaces, extracted from the 
quasi-hole energy dispersion, is consistent with results obtained by ARPES. Furthermore, we 
have calculated the spectral weights (SW's) of the quasi-hole 
and -particle excitations and made comparison of the quantities derived from them with 
the observation made by STM/STS.  

%to 
%compare with results by other numerical approaches (see \cite{lhn,llhl} 
%and references therein) and experiments. 

%While results
%agree fairly well with experiments, these studies on 
%doping holes and electrons into the system 
%only emphasize the {\em asymmetry} resulting from the different signs of long-range 
%hopping amplitudes $t'$ and $t''$ for the corresponding 
%Hamiltonians. It is unclear whether the same physics is working for these 
%two systems with {\em different} Hamiltonians and small Fermi surfaces 
%and quasi-particle states emerge.
%Specific trial wavefunctions (TWF's) describing well the low-energy states 
%of the associated $t$-$t'$-$t''$-$J$ model Hamiltonian with 
%lightly and overdoped 
%doped holes and/or electrons are found to behave also similarly with that in real materials. 

{\bf \em Evolution of ground states with doping.} Let us first discuss the TWF's 
for ground states of the $t$-$J$-type Hamiltonian {\bf \em H}=$\sum_{i,j \sigma}-{t_{ij}}$ 
$\tilde{c}^{\dagger}_{i\sigma}\tilde{c}_{j\sigma}
%-t'\sum_{\langle i,l\rangle \sigma}$ 
%$\tilde{c}^{\dagger}
%_{i\sigma}\tilde{c}_{l\sigma}-t''\sum_{\langle i,m\rangle \sigma}$ 
%$\tilde{c}^{\dagger}
%_{i\sigma}\tilde{c}_{m\sigma}$
+H.c.$+{\bf \em H}$_J$ with different numbers of holes or electrons. Here 
$i,j $ in the summation represents nearest-neighbor ($n.n.$) sites the electron 
hops (with amplitude $t$), the second $n.n.$ ($t'$), third 
$n.n.$ site ($t''$) pairs and the Heisenberg exchange term 
{\bf \em H}$_{J}$=$J\sum_{\langle i,j \rangle}({\bf S}_i\cdot{\bf S}_j-{1\over4} 
n_in_j)$ between $n.n.$ electrons. Note that $\tilde{c}_{i\sigma}$ in {\bf \em H} 
creates different kinds of {\em holes} from 
single-electron-occupied sites: 
{\em empty holes} (0{\it e}-hole) for hole doping 
and {\em two-electron-occupied holes} (2{\it e}-hole) for electron 
doping \cite{clarke}. Namely, $\tilde{c}_{i\sigma}$ is actually equal to 
$c_{i\sigma}(1-n_{i,-\sigma})$ or $c_{i,-\sigma} n_{i\sigma}$ for hole or 
electron doped case, respectively.

Taking the point of view of doping the Mott insulator with antiferromagnetic long range 
order (AF LRO), we discuss TWF's determined variationally at different 
doping levels from the undoped case. The TWF's we constructed may include three
mean-field parameters, depending on the doping concentration: the staggered magnetization
$m_s$=$\langle S^z_A\rangle$=$-\langle S^z_B\rangle$, where the lattice
is divided into two sublattices when AF LRO is present; the uniform bond 
order parameters 
$\chi$=$\langle \sum_{\sigma}c^{\dagger}_{i\sigma}c_{j\sigma}\rangle$; 
and $d$-RVB order
$\Delta$=$\langle c_{j\downarrow}c_{i\uparrow}-c_{j\uparrow}c_{i\downarrow}
\rangle$ if $i$ and $j$ are $n.n.$ sites in $x$-direction and $-\Delta$ in $y$-direction. 

The TWF for lightly doped case is a generalization of 
the single-hole WF with explicit AF LRO first written down by 
Lee and Shih \cite{lee-shih}. In contrast to states at higher doping (see below), WF in this 
regime is constructed solely from the optimized one at half-filling, {\it i.e.} 
with Hamiltonian {\bf \em H}$_J$, 
\begin{equation}
|\Psi_0\rangle = P_d [{\sum_{\bf k [SBZ]} (A_{\bf k} 
a^{\dagger}_{{\bf k}\uparrow}a^{\dagger}_{{\bf -k}\downarrow}+B_{\bf k}
b^{\dagger}_{{\bf k}\uparrow}b^{\dagger}_{{\bf -k}\downarrow})}]^{N_s/2} | 0 \rangle .\\ 
\label{e:undoped} 
\end{equation}
Here coefficients 
$A(B)_{\bf k}$
=$[+(-)E_{\bf k}+\xi_{\bf k}]/\Delta_{\bf k}$ 
%and 
%$B_{\bf k}$
%=$-(E_{\bf k}-\xi_{\bf k})/\Delta_{\bf k}$ 
%are functions of $\xi_{\bf k}$ and $\Delta_{\bf k}$.
with $E_{\bf k}$=($\xi_{{\bf k},0}^2+\Delta_{\bf k}^2$)$^{1/2}$. 
$\xi_{{\bf k},0}$=
$[\epsilon_{\bf k}^2+(m_{sv})^2]^{1\over2}$ 
are energy dispersions for the two spin density wave (SDW) bands 
with 
$\epsilon_{\bf k}$=$-(\cos{\rm k}_x+\cos{\rm k}_y)$. 
%and the staggered magnetization
%$m_s$=$\langle S^z_A\rangle$=$-\langle S^z_B\rangle$, where the lattice 
%is divided into A and B sublattices. 
$a_{{\bf k}\sigma}$ and $b_{{\bf k}\sigma}$ represent the operators 
of the lower and upper SDW bands, respectively,  and are 
related to the original electron operators $c_{{\bf k}\sigma}$ and 
$c_{{\bf k}+{\bf Q}\sigma}$ with  ${\bf Q}$=$(\pi,\pi)$ set for the 
commensurate SDW state. $\Delta_{\bf k}$=$\Delta_{v}(\cos{k_x}-\cos{k_y})$ 
for the $d$-RVB order parameter.  
The projection operator $P_d$
enforces the constraint of 
%no doubly occupied (or vacant) sites for cases 
%with finite hole (or electron) doping . 
one electron per site.
At half-filling, $N_s$- number of lattice sites- equals the total number of electrons. 
Notice that the sum in $|\Psi_{0} \rangle$ is taken over sublattice BZ (SBZ). 
There are two variational
parameters: $\Delta_{v}$ and $m_{sv}$ in this WF.
It has been shown that this WF produces both the staggered magnetization, 
$\langle m \rangle$=${N_s}^{-1}{\sum}_{i}(-1)^{i}S_{i}^{z}$, and optimized energy 
very similar to those obtained by quantum Monte Carlo approach \cite{liang}.  

%we also 
%found that the 
%preference of {\bf Q}/2 for holes 
%causes clearly larger disturbance of the AF order than for the electron 
%doped case with momentum $(\pi,0)$ 
%%shows much less influence on the AF order 
%\cite{lhn}. This is consistent with
%previous ED work and also the  
%experimental results that
%AF phase is more stable for electron doping than hole doped case \cite{takagi}.

When a hole is doped or an electron is removed from the Mott "vacuum" 
$|\Psi_{0} \rangle$, a pair must be broken with an unpaired spin left. Thus it 
is quite natural to have the following Lee-Shih WF for a single doped hole and 
a lone up spin, for example,
\begin{eqnarray}
|\Psi_1({\bf q})\rangle & = &
P_d~c^{\dagger}_{{\bf q}\uparrow} \nonumber \\
& & \mbox{} [{\sum_{[{\bf k} \neq {\bf q}]} 
(A_{\bf k} a^{\dagger}_{{\bf k}\uparrow}a^{\dagger}_{{\bf -k}\downarrow}
 +B_{\bf k} b^{\dagger}_{{\bf k}\uparrow}b^{\dagger}_{{\bf -k}\downarrow})}
]^{(N_s/2)-1}
|0\rangle ,  \nonumber
\label{twf-sb}
\end{eqnarray}
where the hole momentum ${\bf q}$ is excluded from the sum
if ${\bf q}$ is within the SBZ, otherwise, ${\bf q}-{\bf Q}$ is excluded. 
%$P_d$ here enforces the constraint of no doubly occupied sites. 
%When we choose 
%the unpaired-spin momentum ${\bf q}$ to be either
%the same as the hole momentum ${\bf q}_{h}$ or ${\bf q}_{h}$+${\bf Q}$,
%his WF is equivalent to the Lee-Shih one \cite{lee-shih}. 
Note that this WF 
does not contain any information about hoppings of the doped hole or electron. 
Nevertheless, the effect of $n.n.$ hopping is included in the RVB 
uniform bond which describes the large quantum fluctuation and spin singlet 
formation. There is also no need to introduce $t'$ and $t''$ in the TWF as they
are compatible with AF LRO.
Variational energies calculated vary with  ${\bf q}$ and we may obtain 
the energy dispersion for the $t$-$t'$-$t''$-$J$ model \cite{lhn,lee-shih}. 
%The energy dispersions for $t$-$J$ and $t$-$t'$-$t''$-$J$ models are plotted 
%as filled circles in Fig. 1(a) and (b), respectively. 
%For both models, 
The ground state with one hole is found to have momentum $(\pi/2,\pi/2)\equiv {\bf Q}/2$. 
As shown in Ref.\cite{lhn},
this dispersion relation is still followed when the hole number is increased.  
%The holes in these WF's behave just like QP's, hence
%we denote $|\Psi_1({\bf q}_{h}$=${\bf q}_{s})\rangle$$\equiv$$|\Psi_1^{QP}\rangle$.
It is interesting to note that states in the hole- and electron-doped cases are in 
one-to-one correspondence after 
a local particle-hole transformation $c_{i\sigma}\rightarrow c_{i,-\sigma}^{\dagger}$ is made. 
Also, because of the Fermi statistics, 
the exchange of a single spin with a 2{\it e}-hole has an 
extra {\em minus} sign as compared to the 0{\it e}-hole. 
Hence, the only difference between the hole and 
electron doped $t$-$t'$-$t''$-$J$ models is $t'/t \rightarrow -t'/t$ and 
$t''/t \rightarrow-t''/t$
after we change the $c_{i\sigma}$ on one sublattice sites to $-c_{i\sigma}$ \cite{toh-mae94}. 
With all these, we treat the hole and electron doped cases in the 
same manner. As for the case of having an extra up-spin electron with momentum ${\bf q}$ 
doped into the half-filled state, the energy dispersion can be calculated 
with this same $|\Psi_{1} \rangle$. The ground state is here at momentum $(\pi,0)$. 
This result agrees well with that obtained by self-consistent Born approximation (see 
the discussion in Ref.\cite{lhn}). 
Dispersions for single-hole and -electron cases turn out to be simply the combination of the 
mean-field band at half-filling and the coherent hoppings \cite{lee-shih,lhn}.
Using $|\Psi_{1} \rangle$, we calculated the momentum 
distribution function (MDF) 
%$\langle n_{\sigma}^{h}({\bf k})\rangle$ 
$n_{\bf k}=\langle c^{\dagger}_{{\bf k}\sigma}c_{{\bf k}\sigma} \rangle$ for the ground state 
of, say, a single 0$e$-hole with momentum {\bf Q}/2.
% and  $S_z$=$1/2$.
%Results are shown in Fig.{\ref{nk1-eh}}(a) and (b). 
We observed the same  
dips or pockets at {\bf Q}/2 and {\em anti-dips} at -{\bf Q}/2 
found for the exact results of 32 sites (see Ref. \cite{lhn}).
% are also clearly seen here.
%$\langle n_{\sigma}^{e}({\bf k})\rangle$ 
%for electron doped systems could be also calculated from  $|\Psi_{1}\rangle$
%if we perform the transformation, $c_{i\sigma}\rightarrow c_{i,-\sigma}^{\dagger}$, 
%and $t$ is chosen to be positive.
%It is quite amazing that $|\Psi_{1} \rangle$, including no 
%$t'$ and $t''$, not only produces the correct energy dispersions 
%for a single doped hole or electron it also provides a correct picture
%about the momentum distribution. 
%with the Lee-Shih WF originally proposed only for a single hole. 
%The VMC results presented in this paper are for
%$J/t=0.3$, $t'/t=-(+)0.3$ and $t''/t=+(-)0.2$ in the hole(electron) 
%doped case following the values usually used \cite{toh-mae94}.  
%In the following, we concentrate only on the hole doped case.

%%%%%%%%%%%%%%%%%% 1.phase-diagram  %%%%%%%%%%%%%%%%%%

\begin{figure}[top]
\rotatebox{0}{\includegraphics[height=2.2in,width=3.0in]{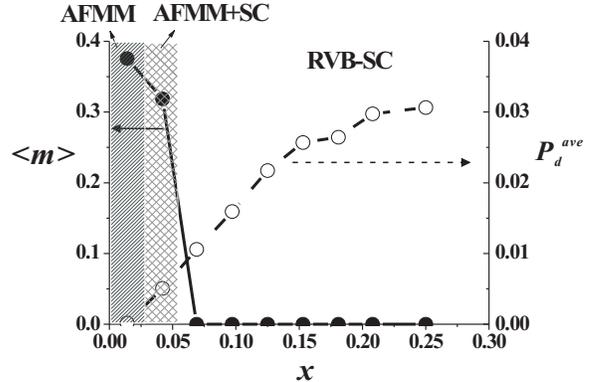}}
\caption{Ground state phase diagram for the $t$-$t'$-$t''$-$J$ model with 
($t'$,$t''$)/$t$=($-0.2,0.1$). Shaded and crossed-line regions represent 
schematic doping regions of AFMM and 
coexisting AFMM+SC states, respectively. $\langle m \rangle$ represents 
the staggered magnetization value (vertical axis on the left) and $P^{ave}_d$ (
axis on the right) for SC pairing 
amplitude (see text for their definition). The results are obtained on the 
lattice of size $12\times12$.}
\label{phase-diagram}
\end{figure}

%%%%%%%%%%%%%%%%%% 1.phase-diagram  %%%%%%%%%%%%%%%%%%

We may generalize the WF to describe states with two doped holes by taking out the
unpaired spin from  $|\Psi_{1} \rangle$. The lowest energy state turns out to be the one 
with zero total momentum and $S_z$=$0$. Namely, the TWF for two holes with momenta  
${\bf q}$ and  $-{\bf q}$ is 
\begin{equation}
|\Psi_2\rangle = P_d[{\sum_{\bf k} {}' 
(A_{\bf k} a^{\dagger}_{{\bf k}\uparrow}
a^{\dagger}_{{\bf -k}\downarrow}
 +B_{\bf k} b^{\dagger}_{{\bf k}\uparrow}b^{\dagger}_{{\bf -k}\downarrow})}
]^{(N_s/2)-1} |0\rangle .  \\ 
\label{e:AFMM}
\end{equation}
Note that momentum ${\bf q}$ is not included in the summation.  
It is most surprising to find that, although $|\Psi_2\rangle$ has 
zero total momentum irrespective of ${\bf q}$,
its energy varies with the missing momentum or 
{\it the hole momentum} ${\bf q}$. Same type of TWF for four holes
with momenta $\pm {\bf q}$ and $\pm {\bf q'}$ both excluded from the 
summation can be constructed. 
%The values of the two parameters $\Delta/\chi$ and $m_{s}/\chi$ 
%are the same for $|\Psi_2\rangle$ and $|\Psi_1\rangle$. Similarly, 
The energy dispersion for two holes doped into the 
half-filled state 
%is shown in  Fig.{\ref{2eh-disp}}(b). Again its dispersion
is also almost identical to that of a single hole and the 
minimum is at {\bf Q}/2 \cite{lhn}. 
%The lowest energy obtained is $-26.438(3)t$ for 8$\times$8 system 
%(... for 4 holes in 14$\times$14 one) 
%which is much lower than the variational energy,
%$-25.72(1)t$ [$-25.763(7)t$ when $t'$ and $t''$ are included], using the 
%TWF, say, applied by Himeda and Ogata \cite{himeda-ogata}.
%Even if we include  $t'$ and $t''$, the variational energy 
%$-25.763(7)t$ (... for 4 holes in 14$\times$14 one), is still much higher than ours \cite{lch}.
%In the inset of  Fig.{\ref{2eh-disp}}
%We also found that 
%the hopping amplitudes for $n.n.$, second $n.n.$ and third $n.n.$ 
%are shown for one hole and two holes as a function of  ${\bf q}$. 
%It shows that the values 
%of two holes are almost twice that of one hole.
The MDF for this state has both dips at
{\bf Q}/2 and -{\bf Q}/2. This is in good agreement with
the exact result for the $t$-$t'$-$t''$-$J$ model with 2 holes 
in 32 sites. For 2 electron doped case, the dispersion 
%shown in Fig.{\ref{2eh-disp}}(a) is 
turns out to be very similar to
that of a single doped electron \cite{lhn}. 
The state with momentum $(\pi,0)$ has the lowest energy for two 
electrons. 

Our results reproduce the contrasting behaviors between lightly hole 
%$Ca_{2-z}Na_{z}CuO_2Cl_2$ ($Na$-CCOC) 
and electron doped [{\em e.g.} $Nd_{2-x}Ce_xCuO_2$ (NCCO)] 
high $T_c$ cuprates revealed by high-resolution ARPES \cite{arpes,pocket}. 
%Although ARPES on the undoped ({\em i.e.} $z$=$x$=0) 
%insulating 
%state shows an identical energy dispersion of a single hole created below the 
%charge gap, results at a little higher dopings are demonstrated to be 
%different: 
While a small hole patch is observed to be at the center in the nodal 
direction of the BZ [$i.e.$ ({\bf Q}/2)] in hole doped systems,
%in the $Na$-CCOC at even 10 percents doping
small electron patches centered at ($\pi$,0) and (0,$\pi$) are observed in 
lightly doped NCCO.

\begin{figure}[top]
\rotatebox{0}{\includegraphics[height=2.8in,width=3.0in]{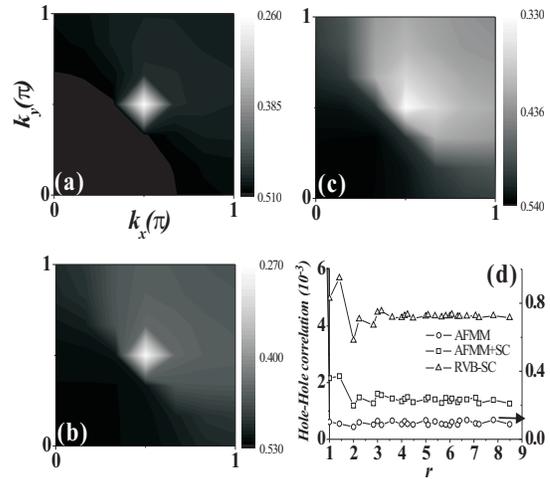}}
\caption{Plot of MDF's for
$12\times12$ sites obtained by (a) AFMM ($x$=0.014), (b) AFMM+SC
($x$=0.042), and (c) RVB-SC ($x$=0.069) states with the same parameters
as that in obtaining results shown in Fig.1. (d) The hole-hole correlation 
functions for different WF's, as denoted by different symbols. Vertical 
axis on the right (for AFMM) is with a finer scale than that on the left.  
}
\label{nk-hh}
\end{figure}

%Applying the same type of TWF for four holes
%with momenta $\pm {\bf q}_{s}$ and $\pm {\bf q'}_{s}$:
%\begin{eqnarray}
%|\Psi_4\rangle &=& P_d~[{\sum_{\bf k} {}' (A_{\bf k} a^{\dagger}_{{\bf k}\uparrow}a^{\dagger}_{
%{\bf -k}\downarrow} 
%+B_{\bf k} b^{\dagger}_{{\bf k}\uparrow}b^{\dagger}_{{\bf -k}\downarrow})}]^{(N_s/2)-2} |0\rangle , 
%\nonumber 
%\label{twf3}
%\end{eqnarray}
%where ${\bf q'}_{s}$ and ${\bf q}_{s}$ are excluded from the summation, 
Focusing on the hole doped case, we found that this type of TWF with even number of holes have 
AF LRO but very little superconducting pairing correlations- namely, $P^{ave}_d$, the averaged 
value of the long-range part of $d$-wave pair-pair correlation function,  
$P_{d}({\bf r})$=$(N_s)^{-1}\sum_{i} \langle \Delta_{i}^{\dagger}\Delta_{i+{\bf r}} \rangle$ 
where $\Delta_{i}$=$c_{i,\uparrow}(c_{i+{\bf x},\downarrow}+
c_{i-{\bf x},\downarrow}-c_{i+{\bf y},\downarrow}-c_{i-{\bf y},\downarrow})$, for $|{\bf r}|>2$ 
in the $12 \times 12$ 
lattice is only about $10^{-4}$ ({\bf x} and {\bf y} are unite vectors in the {\it x} and {\it y} directions, 
respectively). The tiny pairing strength is also indicated by the very small value of 
$n.n.$ hole-hole correlation function, $(N_{s})^{-1} \sum_{i}\langle(1-n_{i})
(1-n_{i+{\bf r}})\rangle$ with $n_{i}$ the electron number at site $i$, as shown in Fig. {\ref{nk-hh}}(d). 
This state with large staggered 
magnetization, small Fermi patch, but no SC pairing behaves like an AFMM [see 
Fig. {\ref{phase-diagram}} and Fig. {\ref{nk-hh}}(a) and (d)] \cite{lhn}. Note that this state 
has been found in the inner 
planes of the Hg-based 5-layered cuprates $HgBa_{2}Ca_{4}Cu_{5}O_{12+\delta}$ \cite{NMR}. 

As more holes are doped into the system, the ground state may be switched from AFMM to the one 
described by WF, 
\begin{equation}
| N_{e}\rangle_{AFMM+SC} =  \nonumber  \\ 
 P_d~[{\sum_{{\bf k}[{\bf SBZ}]} 
(A_{\bf k} a^{\dagger}_{{\bf k}\uparrow}a^{\dagger}_{
{\bf -k}\downarrow} +B_{\bf k} b^{\dagger}_{{\bf k}\uparrow}b^{\dagger}_{{\bf -k}\downarrow})}]^{N_{e}/2} 
|0\rangle,    
\label{e:AFMM-SC}
\end{equation} 
where $N_{e}$ is now the number of electrons and $\pm \xi_{\bf k} \equiv \pm \xi_{{\bf k},1}$=
$\pm [(\cos{\rm k}_x+\cos{\rm k}_y)^{2}+(m_{sv})^2]^{1\over2}-\mu_{v}-
t^{'}_{v}\cos{\rm k}_{x}\cos{\rm k}_{y}-t^{''}_{v}(\cos{2\rm k}_x+\cos{2\rm k}_y)$ containing 3 new variational 
parameters $\mu_{v}$, for chemical potential, and ($t^{'}_{v}$,$t^{''}_{v}$), for long range hoppings. This state, 
which we denote as AFMM+SC, has stronger superconductivity than that of 
AFMM state [$P^{ave}_d$ is about $5\times10^{-3}$, see 
Fig. {\ref{phase-diagram}}, and finite correlation between $n.n.$ holes, see Fig. {\ref{nk-hh}}(d)] and more 
extended MDF [Fig. {\ref{nk-hh}}(b)]. SC and AF ordered states are uniformly mixed or coexisted in the AFMM+SC.   
%diminishes non-linearly (faster than $x^{2}$) from  $10^{-2}$, $5\times10^{-3}$ 
%to $10^{-4}$ with decreasing hole concentration from $10\%$, $4\%$ to $1\%$. 
 
Further increasing the number of doped holes in the system, the ground state becomes the RVB-SC state, namely, 
\begin{eqnarray}
&| N_e\rangle & \equiv P_d | N_e\rangle_{0} = P_d\left(\sum_{{\bf
k}}a_{\bf k} c^\dagger_{{\bf k}\uparrow}c^\dagger_{-{\bf
k}\downarrow}\right)^{N_e/2}\mid 0\rangle,  
\label{e:rvb}
\end{eqnarray}
where $a_{\bf k}=v_{{\bf k}}/u_{{\bf k}}=(E_{\bf
k}-\epsilon_{\bf k})/\Delta_{\bf k}$ with $v_{{\bf k}}$ and
$u_{{\bf k}}$ $d$-wave SC coherent factors. 
Here, $\epsilon_{{\bf k}}=\xi_{{\bf k},1}(m_{sv}=0)$ in the unit of $n.n.$ hopping amplitude.
%-(\cos
%{\rm k}_x+\cos {\rm k}_y)- \mu_{\textit{v}}-t_{\textit{v}}'\cos {\rm k}_x\cos
%{\rm k}_y-t_{\textit{v}}''(\cos 2{\rm k}_x+\cos 2{\rm k}_y)$, 
%$\Delta_{\bf k}=\Delta_{\textit{v}}(\cos {\bf k}_x-\cos{\bf k}_y)$, and $E_{\bf
%k}=\sqrt{\epsilon^{2}_{\bf k} + \Delta^2_{\bf k}}$. 
As shown in Fig. {\ref{phase-diagram}} and Fig. {\ref{nk-hh}}(c) and (d), this state shows even 
stronger SC pairing amplitude and has a large Fermi surface \cite{ltp}. 

The evolution of states, from AF LRO, AFMM, 
AFMM+SC to RVB-SC, we described above for ideal 2-dimensional 
system have been realized recently in multilayer system which is considered to have 
perfectly flat $CuO_2$ planes \cite{NMR}. 
Indeed, the 
ground state phase diagram, Fig. {\ref{phase-diagram}}, as determined for proper set of parameters 
($t',t''$)/$t$ is similar to the one by experiments.

%%%%%%%%%%%%%%%%%% 2.nk+hole-hole correlation  %%%%%%%%%%%%%%%%%%

{\bf \em Properties of the quasi-hole and -particle excitations.} 
We then discuss our attempt to apply the TWF approach to study the properties of 
quasi-hole and -particle excitations of 
the RVB-SC ground state. We have calculated the quasi-hole excitation spectrum of 
the SC state and deduced the normal-state Fermi surface. Without any disorder in 
the system, we also examined the  
effects of strong correlation exactly by investigating the SW's of the uniform 
SC state on finite square lattices \cite{condmat}.  

%%%%%%%%%%%%%%%%%%

\begin{figure}[top]
\rotatebox{0}{\includegraphics[height=3.2in,width=3.0in]{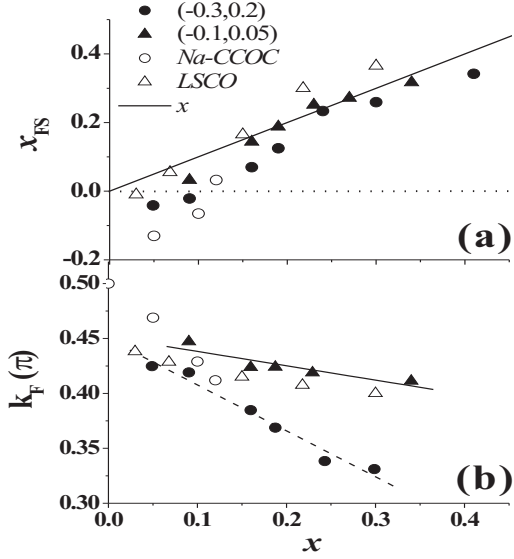}}
\caption{Doping dependence of (a) the hole density $x_{FS}$ and (b) the ${\bf k}_{F}$ 
position at the nodal point obtained from the fitted quasi-hole dispersion. (See 
text for the detail.)
Solid line in (a) represents the relation $x_{\bf FS}=x$. Solid circles 
and triangles are numerical results for
$(t',t'')/t$=(-0.3,0.2) and (-0.1,0.05), respectively.
Dashed and solid lines in (b) are linear fitting to these two sets of results. 
Empty symbols represent experimental data extracted from ARPES, with 
$Na-CCOC$ for $Ca_{2-x}Na_{x}CuO_2Cl_2$ and $LSCO$ for $La_{2-x}Sr_{x}CuO_{2}$ 
\cite{yoshida}. 
}
\label{fig-3}
\end{figure}

We consider the quasi-particle excitation defined by 
\begin{equation}
\mid N_e+1\rangle \equiv P_dc^{\dagger}_{{\bf k}\sigma}\mid
{N_e}\rangle_{0},
\end{equation}
and the quasi-hole one, 
\begin{equation}
\mid N_e-1\rangle \equiv P_dc^{\dagger}_{-{\bf k}-\sigma}\mid
{N_e-2}\rangle_{0}. 
\label{e:h-exc}
\end{equation}
By applying the $t$-$t'$-$t''$-$J$ Hamiltonian to $| N_e-1\rangle$,  
we calculate its excitation energy for each momentum and then fit 
the results by the excitation dispersion, $E_{\bf k}$=
($\epsilon_{{\bf k}}^2+\Delta_{\bf k}^2$)$^{1/2}$
with $\epsilon_{{\bf k}}=\xi_{{\bf k},1}(m_{sv}=0)$. The Fermi surface is then 
determined by the zero energy contour of $\epsilon_{{\bf k}}$. The doping 
dependence of the hole density, deduced from the area of the determined 
Fermi surface is shown in Fig. {\ref{fig-3}}(a). With different sets of ($t',t''$)/$t$ 
which are believed to be material dependent- for example, in Fig. {\ref{fig-3}}(a), 
(-0.3,0.2) for $Ca_{2-x}Na_{x}CuO_2Cl_2$ 
and (-0.1,0.05) for $La_{2-x}Sr_{x}CuO_{2}$ \cite{lmto,yoshida,kyyang}, the results 
are consistent with 
that determined by ARPES. We may also examine the Fermi surface position along the 
nodal direction, {\it i.e.} along (0,0)-($\pi,\pi$), at each hole doping 
density [see Fig. {\ref{fig-3}}(b)]. 
As can be seen clearly, the Fermi surface position moves more prominently for the 
case with larger ($t',t''$)/$t$ values, again consistent with the 
experimental observation (see also Ref.\cite{tanaka} for Bi-based bilayer cuprates).    

%%%%%%%%%%%%%%%%%%%%%%%%%%%%%%%%%% 3.FS %%%%%%%%%%%%%%%%%%%%%%%%%%%%%%%%%%%%%%%

%%%%%%%%%%%%%%%%%%%%%%%%%%%%%%%%%%% 3.FS %%%%%%%%%%%%%%%%%%%%%%%%%%%%%%%%%%%%%

%%%%%%%%%%%%%%%%%%%%%%%%%%%%%%%%

Let us then move to calculate the SW for adding (and
removing) one electron defined by 
\begin{eqnarray}
Z_{{\bf k}\sigma}^{+(-)} = \frac{\mid\langle N_e{+(-)}1\mid
c^{\dagger}_{{\bf k}\sigma}(c_{{\bf k}\sigma})\mid
N_e\rangle\mid^2}{\langle N_e\mid N_e\rangle\langle N_e{+(-)}1\mid
N_e{+(-)}1\rangle}, \label{e:sw1}
\end{eqnarray}
%where
%\begin{equation}
%\mid N_e+1\rangle \equiv P_dc^{\dagger}_{{\bf k}\sigma}\mid
%{N_e}\rangle_{0}
%\end{equation}
%for the quasi-particle excitation, and
%\begin{equation}
%\mid N_e-1\rangle \equiv P_dc^{\dagger}_{-{\bf k}-\sigma}\mid
%{N_e-2}\rangle_{0} \label{e:h-exc}
%\end{equation}
%for the quasi-hole one. 
Applying identities for projection operator, 
\begin{eqnarray}
[c_{{\bf k}\sigma},P_d]P_d &=& 0 ;\nonumber\\
P_d c_{{\bf k}\sigma}[c^{\dagger}_{{\bf k}'\sigma},P_d]&=&
P_d[\frac{1}{N_s}\sum_{i}e^{i({\bf k}'-{\bf k})\cdot{\bf R}_
{i\sigma}}n_{i,-\sigma}]P_d
\label{e:identity1}
\end{eqnarray}
with ${\bf R}_{i\sigma}$ the position vector of the $i$-th spin $\sigma$ in 
the lattice of size $N_s$ and $n_{i\sigma}=c^{\dagger}_{i\sigma}c_{i\sigma}$,
we can relate $Z_{\bf k\sigma}^{+}$ exactly to the hole doping and MDF as
\begin{equation}
Z_{{\bf k}\sigma}^{+}=\frac{1+x}{2}-n_{{\bf k}\sigma} \label{e:relation}
\end{equation}
%where {\it x} is the density of doped holes and $n_{\bf k}=\langle N_e\mid
%c^{\dagger}_{{\bf k}\sigma}c_{{\bf k}\sigma}\mid
%N_e\rangle / \langle N_e\mid N_e\rangle$ 
\cite{yunoki,nave,yang}.

On the other hand, reminiscent of what have been argued previously by analytic 
approach \cite{rantner-wen}, we recognize that the strong correlation 
effects becomes apparent only in $Z_{{\bf k}\sigma}^{-}$ at low doping.
%Unfortunately, Eq.(\ref{e:pairing}) conceals this behavior 
%in the combination with the electron adding one. 
The effects due to strong correlation are 
examined by comparing the {\it coherent} SW averaged over all momenta, {\it i.e.} 
$Z^{-}_{ave}\equiv\sum_{{\bf k}}Z_{{\bf k}\sigma}^{-}/N_s$, and the {\it incoherent} part defined by the relation 
\begin{equation}
n_{ave}^{inc}\equiv n_{ave}-Z^{-}_{ave}  
\label{e:incoherent}
\end{equation}
obtained by exact treatment of the projection and by using renormalized mean-field theory (RMFT) \cite{rmft}. Here 
$n_{ave}\equiv\sum_{{\bf k}}n_{{\bf k}\sigma}/N_s$ is the average MDF which should always be equal to the 
electron density of the system.  

\begin{figure}[top]
\rotatebox{0}{\includegraphics[height=3.4in,width=3.0in]{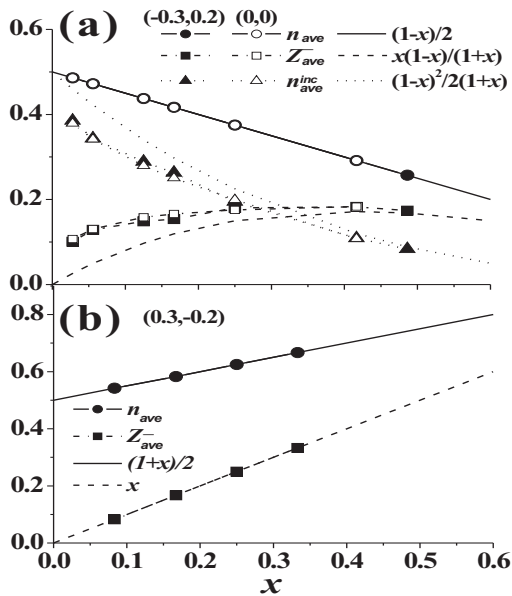}}
\caption{The doping dependence of average SW and MDF of RVB-SC state in hole 
doped (a) and electron doped (b) systems, obtained numerically for 
the $12\times12$ lattice and by RMFT.
Numerical and expected $n_{ave}$ are denoted by the circles and the
solid line, respectively. The squares (triangles), connected by
dashed (dotted) line as the guide for the eye, are for exact results
of $Z^{-}_{ave}$ [$n_{ave}^{inc}$ only for hole doped case,
extracted using Eq.(\ref{e:incoherent}) with $n_{ave}$ calculated
numerically]. The dashed and dotted lines without data points
represent results by RMFT. All solid symbols in (a) are results for
$(t',t'')/t$=(-0.3,0.2) and empty ones for (0,0). Results in (b) are all
for $(t',t'')/t$=(0.3,-0.2).
}
\label{fig-4}
\end{figure}

The exact results for the $12\times12$ lattice and that by RMFT are shown in Fig.\ref{fig-4}(a). 
The coherent part of $Z_{{\bf k}\sigma}^{-}$ by RMFT is 
$g_{t}v^{2}_{\bf k}$ with renormalization factor $g_{t}=2x/(1+x)$. 
Completing the momentum sum for the coherence factor, the average result is $x(1-x)/(1+x)$ and, thus, 
$n_{ave}^{inc}=[(1-x)^{2}]/2(1+x)$ \cite{sum-rule}. The RMFT results are plotted in Fig.\ref{fig-4}(a) (dashed 
and dotted lines) in comparison with the exact ones. As is shown there, while 
the numerical $n_{ave}$ is indeed equal to the electron density, the exact incoherent 
SW for removing an electron is less than the RMFT result. The difference becomes more 
significant as 
hole doping level is reduced. Interestingly, this behavior is independent of the $(t',t'')/t$ 
values [represented by solid and empty symbols in Fig.\ref{fig-4}(a)] which correspond to very 
different doping dependence of the Fermi surface shape 
and also the density of states. By contrast, the average values of $Z_{{\bf k}\sigma}^{+}$ calculated exactly (not 
shown) and by RMFT are identical due to Eq.(\ref{e:relation}). 
%Moreover, they are smaller than $Z^{-}_{ave}$ for the same 
%$12\times12$ lattice as doping density is less than about 0.1 (not shown). The different values of 
%average spectral weights between removing and adding an electron already indicate the 
%particle-hole asymmetric behavior due to strong correlation.     

As a brief digression to electron doped case, it is straightforward
to show that, applying the hole-particle transformation to
Eq.(\ref{e:relation}),
%and thus $Z_{\bf k\sigma}^{+(-)} \rightarrow Z^{-(+)}_{{-{\bf k}+(\pi,\pi)\sigma}$,
SW for removing an electron in electron doped system satisfies the relation 
$Z^{-}_{\bf k\sigma}$=$n_{\bf k\sigma}-(1-x)/2$ rigorously. The doping dependences of
$Z^{-}_{ave}$ and $n_{ave}$ are shown in Fig. 4(b). The numerical
results is exactly equal to RMFT ones indicating that there is no incoherent part of SW 
for removing an electron in the electron doped system.

%%%%%%%%%%%%%%%%%%%%%%%%%% 4.numerical+RMFT %%%%%%%%%%%%%%%%%%%%%%%%%%%%%%

%%%%%%%%%%%%%%%%%%%%%%%%%% 4.numerical+RMFT %%%%%%%%%%%%%%%%%%%%%%%%%%%%%%

\begin{figure}[top]
\rotatebox{0}{\includegraphics[height=2.6in,width=3.0in]{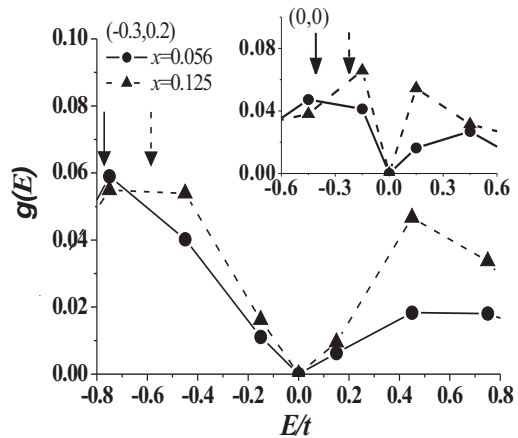}}
\caption{$g(E)$ for RVB-SC state versus excitation energy $E$ for
2 different dopings. Results shown here are for $(t',t'')/t$=(-0.3,0.2)
and lattice size $12\times12$. The associated excitation
gap positions are marked by arrows (see text). \textbf{Inset}: Same
plots for $(t',t'')/t$=(0,0). Results for two doping concentration are denoted 
by the same symbols as that in the main figure.
%$Z^{-}_{\bf k}$ at hole doping $x$ =
%0.07 along the high symmetry directions of {\bf k}, (0,0) $\rightarrow$ ($\pi$,0), ($\pi$,0)
%$\rightarrow$ ($\pi$,$\pi$), and ($\pi$,$\pi$) $\rightarrow$ (0,0). Solid circle and 
%empty square denote RVB-SC and projected Fermi liquid (pFL) WF's, respectively. 
%$\bf{Inset}$: Same plots for $d$-BCS and FL WF's at the same doping level.
}
\label{fig-5}
\end{figure}

To make a comparison with tunneling experiments, we then concentrate on the 
SW's as a function of the excited-state energy. 
%By applying the $t$-$t'$-$t''$-$J$ 
%Hamiltonian to excitations $\mid N_e\pm 1\rangle$, we may extract their 
%excitation energies for each momentum and also the corresponding energy 
%gap by fitting the excitation energy $E_{\bf k}$. 
To reduce the effect of finite size, we define the 
sum of $Z_{{\bf k}\sigma}^{\pm}/N$, over momentum {\bf k} which has energy within $E-{\Delta E}/2$ and 
$E+{\Delta E}/2$, as $g(E)$ 
%[negative(positive) for removing(adding) an electron] 
which could be viewed, approximately, as proportional to the conductance at low energy $E$. We plot 
$g(E)$ in Fig. \ref{fig-5}, up to about the energy where peaks appear for lattices of size $12\times12$ with 
$\Delta E/t=$0.3 for various dopings. (The energy interval we chose is to reduce the 
effects due to lattice size and also to include about the same number of momentum
points below and above zero energy.)
To make sure our treatment is correct, we have also applied the same analysis to 
the {\it d}-wave BCS ($d$-BCS) state and found the ideal BCS result is hardly distorted by the finite size. 
Therefore, with the reasonable finite-size dependence, we obtain indeed the {\it V}-shape 
$d$-wave gap near zero energy. The width between peak positions is also roughly equal to 
two times of the gap value deduced from the excitation energy. Looking at the results closely, while $g(E)$ 
may indeed be about the same at the opposite sides in the very vicinity of zero energy as 
suggested in Ref.\cite{yang}, $g(E)$ for removing an electron is always larger than that for adding an electron 
at higher energy near that of the peak. With decreased doping, the ratio of $g(E)$ at negative and positive energies 
enhances quite dramatically, {\it e.g.} from $x$=0.125 to 0.056, 
$g(-{\Delta})/g({\Delta})$ at the corresponding peak energy $\Delta$ (in units of $t$)
increases from 1.96 to 2.73. Similar behaviors are found for the case with vanishing 
$(t',t'')/t$, as shown in the inset of Fig. \ref{fig-5}. In contrast to this, for the $d$-BCS 
case in the same finite lattices there is almost no change of the ratio 
within the gap \cite{condmat}. The numerical results thus tells us features due to strong correlation, {\it i.e.} 
the particle-hole asymmetry of average conductance exists even within the gap region and gets 
enhanced with underdoping, which are not yet fully explored in the tunneling experiment.    

We can also see, from Fig. \ref{fig-5}, the correlation between heights of the spectral
weight peak and the gap size as doping level is varied. For the two doping level shown 
there, the peak height scales with the
pairing amplitude but anti-correlates with the gap size.
This is in clear contrast to the BCS case in which the peak height,
proportional to the SC coherence, scales with the gap size as more
holes doped into the system. Our result agrees qualitatively with
what has been extracted from STS experiments \cite{stm1}. 
%It should
%be noted that our result indicates that the width between $Z_{{\bf
%k}\sigma}^{+}$ and $Z_{{\bf k}\sigma}^{-}$ for a strongly correlated
%system is determined by the excitation gap in the SC state. This gap
%was shown in Ref.\cite{fuchun} to behave similarly as the pseudo-gap
%seen by ARPES.

%%%%%%%%%%%%%%%%%% 5. conductance %%%%%%%%%%%%%%%%%%%%%%%%%%%%%%%%

%%%%%%%%%%%%%%%%%% 5. conductance %%%%%%%%%%%%%%%%%%%%%%%%%%%%%%%

To summarize, we discussed $d$-RVB based TWF's describing the ground states of 
the $t$-$J$-type models at different hole doping levels. The AFMM, AFMM+SC 
states have been observed recently in multilayer cuprates. 
We have also determined the doping dependence of both the Fermi surface area and 
Fermi momentum along the nodal direction using the fitted quasi-hole dispersion.
With proper parameters for different real materials, the results are consistent 
with the same quantities observed in various high $T_{c}$ cuprates. By studying 
SW's of the quasi-hole and -particle excitations of the SC state, we found, 
even without any disorder, features qualitatively consistent with what 
has been seen in tunneling experiments. An exact relation is obtained for the 
doping dependence of the spectral 
weight of removing an electron from an electron-doped cuprates. This could be 
tested by ARPES. There are also other results which have to be 
confirmed by future experiments.
 
\vspace*{1cm}

We acknowledge discussions with Profs. C.~T. Shih, Y.~C. Chen, M. Ogata, N. Nagaosa, 
T. Tohyama, 
S. Ishihara, V.N. Muthukumar, A. Fujimori, Drs. Y. Yanase and N. Fukushima. We are also
grateful to Profs. S.~H. Pan, H. Takagi, Y. Kitaoka and Drs. T. Hanaguri and H. Mukuda for 
kindly sharing their results and insights about their experiments. 
TKL and CMH are supported by the National 
Science Council in Taiwan with Grant no.94-2112-M-001-003 and 94-2112-M-032-001, 
respectively. Part of the calculations are performed in the IBM P690 and SMP2 in 
the Nation Center for High-performance Computing in Taiwan.


\begin{references}

\bibitem{bonn2006}J. Orenstein and A.~J. Millis, Science {\bf 288}, 468 (2000);
D.~A. Bonn, Nature Physics {\bf 2}, 159 (2006).
\bibitem{arpes}A. Damascelli, Z. Hussain, and Z.-X. Shen, Rev. Mod. Phys. {\bf 75}, 473 (2003);
J.~C. Campuzano, M.~R. Norman, M. Randeria, in {\it Physics of
Superconductors, Vol. II}, ed. K. H. Bennemann and J. B. Ketterson
(Springer, Berlin, 2004), p. 167-273.
%\bibitem{tsuei-kirtley} C.~C. Tsuei and J.~R. Kirtley, Rev. Mod. Phys. {\bf 72}, 969 (2000).
%\bibitem{gossamer}R.~B Laughlin, Phil. Mag. {\bf 86}, 1165 (2006), also as cond-mat/0209269.
\bibitem{stm0}C. Renner and {\O}. Fischer, Phys. Rev. B {\bf 51}, 9208 (1995); A. Matsuda,
S. Sugita and T. Watanabe, Phys. Rev. B {\bf 60}, 1377 (1999); M.
Kugler {\it et al.}, J. Phys. Chem. Solids {\bf 67}, 353 (2006).
\bibitem{stm1}K.~M. Lang {\it et al.}, Nature {\bf 415}, 412 (2002); K. McElroy {\it et al.},
Phys. Rev. Lett. {\bf 94}, 197005 (2005).
\bibitem{fang}A.~C. Fang {\it et al}, Phys. Rev. Lett. {\bf 96}, 017007 (2006).
\bibitem{stm2}T. Hanaguri {\it et al.}, Nature {\bf 430}, 1001 (2004);
T. Hanaguri, private communication (2006).
\bibitem{pwa87}P.~W. Anderson, Science {\bf 235}, 1196 (1987).
\bibitem{lhn}T.~K. Lee, C.-M. Ho and N. Nagaosa, Phys. Rev. Lett. {\bf 90}, 067001 (2003).
\bibitem{ltp}A. Himeda and M. Ogata, Phys. Rev. B {\it 60}, R9935 (1999); 
C.~T. Shih {\it et al.}, Low Temp. Phys. {\bf 31}, 757 (2005).
\bibitem{NMR}H. Mukuda {\it et al.}, Phys. Rev. Lett. {\bf 96}, 087001 (2006).
\bibitem{clarke} D. G. Clarke, Phys. Rev. B {\bf 48}, 7520 (1993).
\bibitem{lee-shih}T.~K. Lee and C.~T. Shih, Phys. Rev. B {\bf 55}, 5983 (1997).
\bibitem{liang}S. Liang, B. Dou{\c{c}}ot and P.~W. Anderson, Phys. Rev. Lett. {\bf 61}, 365 (1988); 
A.~W. Sandvik, Phys. Rev. B {\bf 56}, 11678 (1997).
\bibitem{toh-mae94}T. Tohyama and S. Maekawa, Phys. Rev. B {\bf 49}, 
3596 (1994); R.~J. Gooding, K.~J.~E. Vos and P.~W. Leung, Phys. Rev. B {\bf 50}, 12866 (1994).
\bibitem{pocket}F. Ronning {\em et al.} 
Phys. Rev. B {\bf 67}, 165101 (2003) for hole-doped systems;
N.~P. Armitage {\em et al.}, Phys. Rev. Lett. {\bf 88}, 257001 (2002), for electron-doped ones. 



%\bibitem{fujimori}K. Tanaka, Ph.D. Thesis, University of Tokyo (2004); section 5.3.4. 

%\bibitem{many}A. Nazarenko {\em et al.}, Phys. Rev. B {\bf 51}, 8676 (1995);
%V.~I. Belinicher, A.~L. Chernyshev, and V.~A. Shubin, {\em ibid.} {\bf 54}, 
%14914(1996); 
%R. Eder, Y. Ohta and G.~A. Sawatzky, {\em ibid.} {\bf 55}, R3414 (1997); 
%F. Lema and A.~A. Aligia, {\em ibid.} {\bf 55}, 14092 (1997);
%C. Kim  {\em et al.}, Phys. Rev. Lett. {\bf 80}, 4245 (1998).
%%E. Pavarini {\em et al.}, {\em ibid.} {\bf 87}, 47003 (2001). 
%\bibitem{vmc}C.T. Shih {\it et al.}, Phys. Rev. Lett. {\bf 92}, 227002 (2004).

\bibitem{condmat}C.-P. Chou, T.~K. Lee and C.-M. Ho, Phys. Rev. B (2006) {\it in press}, also 
as cond-mat/0606633.
\bibitem{lmto}E. Pavarini {\em et al.}, Phys. Rev. Lett. {\bf 87}, 47003 (2001).
\bibitem{yoshida}T. Yoshida {\em et al.}, cond-mat/0510608.
\bibitem{kyyang}K.-Y. Yang {\em et al.}, Phys. Rev. B {\bf 73}, 224513 (2006).
\bibitem{tanaka}K. Tanaka, Ph.D. Thesis (University of Tokyo, 2004); section 5.3.4. 
\bibitem{yunoki}S. Yunoki, Phys. Rev. B{\bf 72} 092505 (2005).
\bibitem{nave}C.~P. Nave {\it et al.}, Phys. Rev. B {\bf 73}, 104502 (2006).
%\bibitem{chou}C.~P. Chou {\it et al.}, unpubished.
\bibitem{yang}H.-Y. Yang {\it et al.}, cond-mat/0604488.
\bibitem{rantner-wen}W. Rantner and X.-G. Wen, Phys. Rev. Lett. {\bf 85}, 3692 (2000).
\bibitem{rmft}F.~C. Zhang {\it et al.}, Supercond. Sci. Tecnnol. {\bf 1}, 36 (1988).
\bibitem{sum-rule}M. Randeria {\it et al.}, Phys. Rev. Lett. {\bf 95}, 137001 (2005).

\end{references}
\end{document}